\documentclass[%
 reprint,
superscriptaddress,
 amsmath,amssymb,
aps,
pra,
showpacs
]{revtex4-2}

\usepackage{graphicx}
\usepackage{bm}
\usepackage{amsthm,amsmath,amssymb}
\usepackage{color}
\usepackage{enumitem}
\usepackage{caption}
\usepackage{subcaption}
\usepackage{mathtools}
\usepackage{dcolumn}
\usepackage{hyperref}
\usepackage[ruled]{algorithm2e}

\definecolor{Red}{rgb}{1,0,0}

\def\tr{\operatorname{tr}}

\def\poly{\operatorname{poly}}

\begin{document}

\preprint{APS/123-QED}

\title{Quantum subspace alignment for domain adaptation}

\author{Xi He}
\email{xihe@std.uestc.edu.cn}
\affiliation{Institute of Fundamental and Frontier Sciences, University of Electronic Science and Technology of China}

\begin{abstract}
Domain adaptation (DA) is used for adaptively obtaining labels of an unprocessed data set with a given related, but different labelled data set. Subspace alignment (SA), a representative DA algorithm, attempts to find a linear transformation to align the subspaces of the two different data sets. The classifier trained on the aligned labelled data set can be transferred to the unlabelled data set to predict the target labels. In this paper, two quantum versions of the SA are proposed to implement the DA procedure on quantum devices. One method, the quantum subspace alignment algorithm (QSA), achieves quadratic speedup in the number and dimension of given samples. The other method, the variational quantum subspace alignment algorithm (VQSA), can be implemented on the near term quantum devices through a variational hybrid quantum-classical procedure. The results of the numerical experiments on different types of datasets demonstrate that the VQSA can achieve competitive performance compared with the corresponding classical algorithm. 

\end{abstract}


\maketitle
\section{Introduction}
\label{sec:introduction}
Transfer learning (TL) is a crucial subfield of machine learning. TL aims to accomplish tasks on an unprocessed data set with the known information of a different, but related data set~\cite{pratt1993discriminability}. In the realm of TL, domain adaptation (DA) specifically attempts to predict the labels of the unprocessed data set based on the given labelled data set. DA has been widely used in natural language processing, computer vision and reinforcement learning~\cite{pan2009survey}. In terms of the labels of the given data, DA can be categorized into the semi-supervised DA and the unsupervised DA. The semi-supervised DA refers to a common method where the unprocessed data set has a few labels~\cite{kulis2011you,duan2009domain,saenko2010adapting,daume2006domain}. The unsupervised DA focuses on the task where the unprocessed data set is totally unlabelled~\cite{gong2013connecting,ni2013subspace,shrivastava2014unsupervised}. As one of the most representative unsupervised DA methods, subspace learning assumes that the data distribution of the labelled data set will be similar to that of the unlabelled data set after some transformations. The subspace learning algorithm mainly contains two types including the statistical-based subspace learning~\cite{fernando2013unsupervised,sun2015subspace,sun2016deep} and the manifold-based subspace learning~\cite{gopalan2011domain,gong2012geodesic} in the view of transformation modes. The former mainly aligns the statistical features of the two data sets with a transformation matrix directly. The latter maps the original data to a lower-dimensional manifold and transforms the labelled data to the target unlabelled data sequentially. 

Subspace alignment (SA)~\cite{fernando2013unsupervised} is one of the most concise and efficient statistical-based subspace learning algorithms. After preprocessing the original data sets to extract the corresponding principal components, SA attempts to find a linear transformation matrix to align the subspace of the labelled data set to that of the unlabelled data set. Subsequently, the classifier can be performed on the aligned subspace data to obtain the target labels. Compared with other DA algorithms, SA is efficient in unsupervised DA and easy to implement. However, with the increase of the scale and dimension of the data samples, the algorithmic complexity of the classical SA can be costly.

Compared with classical computation, quantum computation proposes a new computing pattern utilizing the principles of quantum mechanics~\cite{shor1994algorithms,grover1996fast,harrow2009quantum,aaronson2011computational,farhi2018classification}. It can be applied to the field of machine learning to achieve quantum speedup in computational complexity compared with the corresponding classical algorithms~\cite{lloyd2013quantum,lloyd2014quantum,rebentrost2016quantum}. Concretely, for the shallow machine learning, quantum computing techniques can be applied to dealing with supervised learning tasks such as classification~\cite{rebentrost2014quantum,wiebe2018quantum,dang2018image}, data fitting~\cite{wiebe2012quantum,schuld2016prediction} and unsupervised learning tasks including clustering~\cite{aimeur2013quantum}, dimensionality reduction~\cite{cong2016quantum,he2019quantum_qLLE}. For the deep learning, the feedforward neural network~\cite{wan2017quantum, beer2020training} and the generative models such as quantum auto-encoders~\cite{romero2017quantum,lamata2018quantum,khoshaman2018quantum}, quantum Boltzmann machine~\cite{wiebe2014quantum,amin2018quantum} and quantum generative adversarial network~\cite{lloyd2018quantum,dallaire2018quantum,hu2019quantum,benedetti2019adversarial,situ2018quantum,zeng2019learning} can be implemented on quantum devices. Recently quantum computation can be combined with fine-tuning techniques to accomplish TL tasks on variational hybrid classical-quantum neural networks~\cite{mari2019transfer}. In addition, quantum techniques are applied to the field of DA resulting in performance promotion~\cite{he2020quantum}. 

In this paper, two quantum versions of the SA are proposed. One method, the quantum subspace alignment algorithm (QSA) can be implemented on a universal quantum computer with quadratic speedup in the number and dimension of the given data. In the data preprocessing, we adopt the quantum principal component analysis (qPCA)~\cite{lloyd2014quantum} to transform the given data from the original $D$-dimensional space to their $d$-dimensional subspaces in time $O(d \log D)$ where $d \ll D$. Subsequently, the coordinates of the labelled source domain subspace are aligned to the target domain subspace coordinates by the quantum basic linear algebra subroutines. Compared with the classical SA requiring runtime in $O(D^{2} d)$, the QSA can be implemented on a universal quantum computer in $O(\poly(\sqrt{D}))$. The other method, the variational quantum subspace alignment algorithm (VQSA), can be performed on the near term quantum devices through a variational hybrid quantum-classical procedure. Subsequently, a classifier can be performed on the subspace data to predict the target labels. To evaluate the performance of the VQSA, two different types of numerical experiments utilizing the synthetic data sets and the Iris data set are presented. Based on the parameterized quantum circuits and the classical optimization process, the VQSA can achieve competitive performance compared with the classical SA.

The contents of this paper will be arranged as follows. The classical SA will be briefly reviewed in section~\ref{sec:classical SA}. Based on the classical SA, the QSA will be presented in section~\ref{sec:QSA}. Specifically, the source and target domain data are preprocessed by the qPCA to obtain the corresponding subspace data in section~\ref{subsec:data preprocessing}. Subsequently, the QSA will be implemented in section~\ref{subsec:SA}. We analyze the algorithmic complexity of the classical and quantum SA to demonstrate the superiority of the QSA in section~\ref{subsec:complexity}. The implementation of the VQSA is shown in section~\ref{sec:VQSA}. Two different specific configurations namely the end-to-end VQSA in section~\ref{subsec:E2E_VQSA} and the matrix-multiplication-based VQSA in section~\ref{subsec:MMB_VQSA} are provided. In addition, the numerical experiments are presented in section~\ref{sec:numerical experiments}. Finally, we make a conclusion and discuss some open questions in section~\ref{sec:conclusion}.  

\section{Classical subspace alignment}
\label{sec:classical SA}
\begin{figure}
\centering
\includegraphics[width=0.45\textwidth]{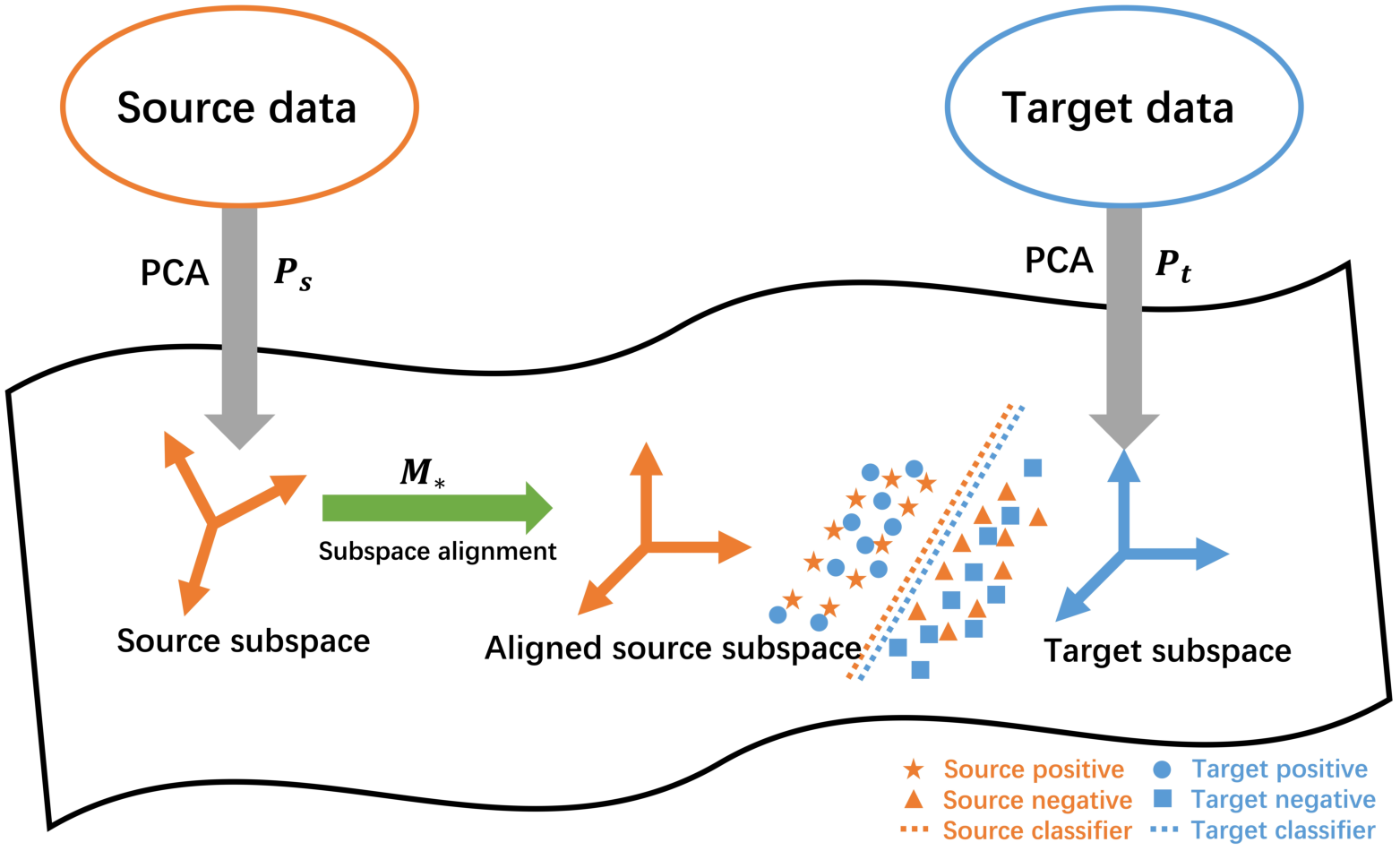}
\caption{The illustration of SA.}
\label{fig:qSA}
\end{figure}

In the field of DA, domain $\mathcal{D}$ refers to the data set $X$ and its corresponding distribution $P(X)$. Given a labelled source domain $\mathcal{D}_{s} = \{ x_{i}^{(s)} \}_{i=1}^{n_{s}} \in \mathbb{R}^{D}$ with labels $Y_{s} = \{ y^{(s)} \} \in \{ 1, \cdots, l \}$, an unlabelled target domain $\mathcal{D}_{t} = \{ x_{j}^{(t)} \}_{j=1}^{n_{t}} \in \mathbb{R}^{D}$, and the corresponding data distributions $P(X_{s}) \neq P(X_{t})$. DA attempts to obtain the labels of $\mathcal{D}_{t}$ by utilizing the knowledge of $\mathcal{D}_{s}$. SA assumes that the source and target domain data depend on a lower-dimensional manifold. After projecting the source domain data $X_{s} = (x_{1}^{(s)}, x_{2}^{(s)}, \cdots, x_{n_{s}}^{(s)}) \in \mathbb{R}^{D \times n_{s}}$ and the target domain data $X_{t} = (x_{1}^{(t)}, x_{2}^{(t)}, \cdots, x_{n_{t}}^{(t)}) \in \mathbb{R}^{D \times n_{t}}$ to their subspaces respectively, SA~\cite{fernando2013unsupervised} aims to align the two subspaces with a linear transformation. So that, a classifier can be applied to the aligned source domain data and the target domain data to predict the target labels $Y_{t} = \{ y^{(t)} \} \in \{1, \cdots, l \}$ of $\mathcal{D}_{t}$. The illustration of the classical SA is presented in Fig.~\ref{fig:qSA}.  

In the first step, by the principal component analysis algorithm (PCA)~\cite{pearson1901liii}, SA projects $X_{s}$, $X_{t}$ to their corresponding $d$-dimensional subspaces $\hat{X}_{s}$, $\hat{X}_{t}$ spanned by the columns of $P_{s}$, $P_{t}$ respectively where $d \ll D$ and $P_{s}, P_{t} \in \mathbb{R}^{D \times d}$.

Subsequently, the source domain subspace coordinates $P_{s}$ can be aligned to the target domain subspace coordinates $P_{t}$ with a transformation matrix $M \in \mathbb{R}^{d \times d}$. Concretely, the corresponding objective function is defined as
\begin{align}\label{eq:objective function}
	M_{\ast} &= \arg \min \limits_{M} \Vert P_{s} M - P_{t} \Vert_{F}^{2} \notag \\
	&= \arg \min \limits_{M} \Vert P_{s}^{T} P_{s} M - P_{s}^{T} P_{t} \Vert_{F}^{2} \notag \\
	&= \arg \min \limits_{M} \Vert M - P_{s}^{T} P_{t} \Vert_{F}^{2},
\end{align}
where $\Vert . \Vert_{F}$ is the Frobenius norm.
Hence, the optimal transformation matrix $M_{\ast} = P_{s}^{T} P_{t}$. The source domain subspace can be aligned to the target domain subspace with $P_{a} = P_{s}M_{\ast}$. The aligned source domain subspace data 
\begin{equation}\label{eq:aligned source domain subspace data}
	\hat{X}_{a} = P_{a}^{T} X_{s} = (P_{s} M_{\ast})^{T} X_{s}
\end{equation}
and the target domain subspace data $\hat{X}_{t}$
\begin{equation}\label{eq:target domain subspace data}
	\hat{X}_{t} = P_{t}^{T} X_{t}
\end{equation}
can be obtained.

The similarity function which measures the discrepancy between the source and target domain is defined as 
\begin{align}\label{eq:similarity function}
	sim(x_{s}, x_{t}) &= (\hat{x}_{a})^{T} \hat{x}_{t} \notag \\
	&= x_{s}^{T} P_{s} M_{\ast} P_{t}^{T} x_{t} \notag \\
	&= x_{s}^{T} A x_{t}
\end{align}
where the target aligned matrix $A = P_{s} M_{\ast} P_{t}^{T}$.

After aligning the subspaces, the classifier can be invoked to obtain the target data labels $Y_{t}$. In general, two different types of classifiers can be utilized to accomplish the target label prediction. The local classifier such as the nearest-neighbor algorithm~\cite{gutin2002traveling} can be performed on $\hat{X}_{a}$, $\hat{X}_{t}$ to predict the target labels $Y_{t}$. In addition, the global classifier such as the support vector machine (SVM)~\cite{cortes1995support} can also be utilized to predict the $Y_{t}$ with the similarity function $sim(x_{s}, x_{t})$.

\section{Quantum subspace alignment}
\label{sec:QSA}
In this section, the QSA will be presented. At first, the source and target domain data will be preprocessed by the qPCA in section~\ref{subsec:data preprocessing}. Then, the source and target domain subspace will be aligned in section~\ref{subsec:SA}. Finally, we analyze the algorithmic complexity of the classical and quantum SA algorithms to demonstrate the superiority of the QSA in section~\ref{subsec:complexity}.  

\subsection{Data preprocessing}
\label{subsec:data preprocessing}
In the data preprocessing section, $X_{s}$, $X_{t}$ can be preprocessed by the qPCA~\cite{lloyd2014quantum} to obtain the corresponding principle components and subspace data. The quantum states corresponding to $X_{s}$, $X_{t}$ are 
\begin{align}\label{eq:data set states}
	| \psi_{X_{s}} \rangle = \sum_{i=1}^{n_{s}} \sum_{m=1}^{D} x_{mi}^{(s)} | i \rangle | m \rangle = \sum_{i=1}^{n_{s}} | i \rangle | x_{i}^{(s)} \rangle, \\
	| \psi_{X_{t}} \rangle = \sum_{j=1}^{n_{t}} \sum_{m=1}^{D} x_{mj}^{(t)} | j \rangle | m \rangle = \sum_{j=1}^{n_{t}} | j \rangle | x_{j}^{(t)} \rangle,
\end{align}
respectively in amplitude encoding where $\sum_{m, i} | x_{mi}^{(s)} |^{2} = \sum_{m, j} | x_{mj}^{(t)} |^{2} = 1$. Thus, the quantum state which is proportional to the covariance matrix $C_{s} = X_{s}X_{s}^{T}$ is 
\begin{align}\label{eq:covariance state}
	\rho_{C_{s}} &= \tr_{i} \{ | \psi_{X_{s}} \rangle \langle \psi_{X_{s}} | \} \notag \\
	&= \sum_{m,m^{'}=1}^{D} \sum_{i=1}^{n_{s}} x_{mi}^{(s)} x_{m^{'}i}^{(s) \ast} | m \rangle \langle m^{'} |
\end{align} 
where $\tr_{i}$ is the partial trace over the $i$ register.

By the trick of Ref.~\cite{lloyd2014quantum}, $e^{i\rho_{C_{s}}\Delta t}$ can be efficiently simulated. Subsequently, the source domain $d$-dimensional subspace principal components $P_{s} = \sum_{i=1}^{d} | u_{i}^{(s)} \rangle \langle i |$ can be obtained by applying the quantum phase estimation (QPE)~\cite{nielsen2010quantum, duan2017quantum}
\begin{align}\label{eq:phase estimation}
	\textbf{U}_{\textbf{PE}}(\rho_{C_{s}}) = &(\textbf{QFT}^{\dagger} \otimes \textbf{I})\left( \sum_{\tau=0}^{T-1} | \tau \rangle \langle \tau | \otimes e^{i\rho_{C_{s}}\tau \Delta t} \right) \notag \\
	&(\textbf{H}^{\otimes n} \otimes \textbf{I})
\end{align}
with $O(\Delta t^{2} / \epsilon)$ copies of the quantum state $\rho_{C_{s}}$ and sampling the eigenvectors $| u_{i}^{(s)} \rangle$ corresponding to the $d$ largest eigenvalues of $\rho_{C_{s}}$ where $\textbf{QFT}^{\dagger}$ represents the inverse quantum Fourier transform and $\epsilon$ is the error coefficient. Similarly, the target domain $d$-dimensional subspace principal components $P_{t} = \sum_{j=1}^{d} | u_{j}^{(t)} \rangle \langle j |$ can also be obtained by the qPCA. 

\subsection{Subspace alignment}
\label{subsec:SA}
As presented in the data preprocessing section, the quantum states representing the source and target principal components are
\begin{align}\label{eq:P_{s} P_{t} states}
	| P_{s} \rangle = \sum_{i=1}^{d} \sum_{p=1}^{d} u_{pi}^{(s)} | i \rangle | p \rangle = \sum_{i=1}^{d} | i \rangle | u_{i}^{(s)} \rangle, \\
	| P_{t} \rangle = \sum_{j=1}^{d} \sum_{p=1}^{d} u_{pj}^{(t)} | j \rangle | p \rangle = \sum_{j=1}^{d} | j \rangle | u_{j}^{(t)} \rangle,
\end{align}
respectively where $\sum_{p, i} \vert u_{pi}^{(s)} \vert^{2} = \sum_{p, j} \vert u_{pj}^{(t)} \vert^{2} = 1$. Subsequently, the quantum states
\begin{align}\label{eq:source subspace covariance matrix}
	\rho_{s} &= \tr_{i} \{| P_{s} \rangle \langle P_{s} | \} \notag \\
	&= \sum_{p, p^{'}=1}^{d} \sum_{i=1}^{d} u_{pi}^{(s)} u_{p^{'}i}^{(s)\ast} | p \rangle \langle p^{'} |  
\end{align}
and
\begin{align}\label{eq:target subspace covariance matrix}
	\rho_{t} &= \tr_{j} \{| P_{t} \rangle \langle P_{t} | \} \notag \\
	&= \sum_{p,p^{'}=1}^{d} \sum_{j=1}^{d} u_{pj}^{(t)} u_{p^{'}j}^{(t)\ast} | p \rangle \langle p^{'} | 
\end{align}
which are proportional to $P_{s} P_{s}^{T}$, $P_{t} P_{t}^{T}$ respectively can be obtained.

Inspired from Ref.~\cite{shao2018quantum}, the preparation procedure of the optimal transformation matrix $M_{\ast} = P_{s}^{T} P_{t}$ is presented as follows.

(1) Prepare the initial state. Given the quantum states $| P_{s} \rangle$ and $| P_{t} \rangle$, assume that the quantum oracle
\begin{equation}\label{eq:oracle}
	\textbf{U}_{\textbf{S}}(P): | i \rangle | 0 \rangle \rightarrow | i \rangle | u_{i} \rangle.
\end{equation} 
 is accessible as presented in~\cite{kerenidis2016quantum}. The initial state
\begin{equation}\label{eq:initial state}
	| \psi_{0} \rangle = \sum_{i,j=1}^{d}|i \rangle^{I_{1}} | j \rangle^{I_{2}} | 0 \rangle^{B} | 0 \rangle^{C_{1}}
\end{equation}
can be prepared with four quantum registers denoted by $I_{1}$, $I_{2}$, $B$ and $C_{1}$. Apply the Hadamard operation $\textbf{H}$ and the controlled unitary $\textbf{U}_{\textbf{P}}(P_{s}, P_{t}) = I^{I_{2}} \otimes | 0 \rangle \langle 0 |^{B} \otimes \textbf{U}_{\textbf{S}}^{I_{1}C_{1}}(P_{s}) + I^{I_{1}} \otimes | 1 \rangle \langle 1 |^{B} \otimes \textbf{U}_{\textbf{S}}^{I_{2}C_{1}}(P_{t})$ on the initial state resulting in
\begin{align}\label{eq:psi_1}
	| \psi_{1} \rangle &= \frac{1}{\sqrt{2}}(| 0 \rangle | P_{s} \rangle + | 1 \rangle | P_{t} \rangle) \notag \\
	&= \sum_{i,j=1}^{d} | i \rangle^{I_{1}} | j \rangle^{I_{2}} \otimes | \phi_{0} \rangle^{BC_{1}},
\end{align}
where $| \phi_{0} \rangle = \frac{1}{\sqrt{2}}(| 0 \rangle | u_{i}^{(s)} \rangle + | 1 \rangle | u_{j}^{(t)} \rangle)$.

(2) The quantum state
\begin{align}\label{eq:psi_2}
	| \psi_{2} \rangle &= \sum_{i,j=1}^{d} | i \rangle^{I_{1}} | j \rangle^{I_{2}} \otimes | \phi_{1} \rangle^{BC_{1}}
\end{align}
can be obtained by applying the Hadamard operation $\textbf{H}$ on the $| \phi_{0} \rangle$ register where $| \phi_{1} \rangle = \frac{1}{2} (| 0 \rangle (| u_{i}^{(s)} \rangle + | u_{j}^{(t)} \rangle) + | 1 \rangle (| u_{i}^{(s)} \rangle - | u_{j}^{(t)} \rangle ) )$. Let
\begin{align}\label{eq:phi_1}
	| \phi_{1} \rangle &= \sin \theta_{ij} | 0 \rangle | u_{1} \rangle + \cos \theta_{ij} | 1 \rangle | u_{2} \rangle \notag \\
	&= -\frac{i}{\sqrt{2}} (e^{i \theta_{ij}} | w_{1} \rangle - e^{-i \theta_{ij}} | w_{2} \rangle)
\end{align}
where $\sin \theta_{ij} = \sqrt{(1 + \langle u_{i}^{(s)} | u_{j}^{(t)} \rangle) / 2}$, $\cos \theta_{ij} = \sqrt{(1 - \langle u_{i}^{(s)} | u_{j}^{(t)} \rangle) / 2}$, and $| u_{1} \rangle$, $| u_{2} \rangle$ represent the normalization of $| u_{i}^{(s)} \rangle + | u_{j}^{(t)} \rangle$, $| u_{i}^{(s)} \rangle - | u_{j}^{(t)} \rangle$ respectively. In addition, the quantum states
\begin{equation}\label{eq:w_1, w_2}
	\begin{cases}
		| w_{1} \rangle = \frac{1}{\sqrt{2}} (| 0 \rangle | u_1 \rangle + i| 1 \rangle | u_{2}\rangle), \\
		| w_{2} \rangle = \frac{1}{\sqrt{2}} (| 0 \rangle | u_{1} \rangle - i| 1 \rangle | u_{2} \rangle);	
	\end{cases}
\end{equation}
are the eigenvectors of the matrix $G = (2 | \phi_{1} \rangle \langle \phi_{1} | - I)(-\sigma_{z} \otimes I)$ corresponding to the eigenvalues $e^{\pm i 2 \theta_{ij}}$ where $\sigma_{z}$ is the Pauli-Z operator. 

(3) The QPE $\textbf{U}_{\textbf{PE}}(G)$ is applied on the $| \phi_{1} \rangle$ register resulting in
\begin{align}\label{psi_3}
	| \psi_{3} \rangle &= - \frac{i}{\sqrt{2}} \sum_{i,j=1}^{d} | i \rangle^{I_{1}} | j \rangle^{I_{2}}\otimes (e^{i \theta_{ij}} | w_{1} \rangle^{BC_{1}} | \lambda \rangle^{C_{2}}  \notag \\ 
	&\quad - e^{-i \theta_{ij}} | w_{2} \rangle^{BC_{1}} | - \lambda \rangle^{C_{2}}),
\end{align} 
where $\theta_{ij} = \lambda \pi / 2^{n}$.

(4) Add a new register $R$ and perform the controlled $R_{y}(2\arcsin(\langle u_{i}^{(s)} | u_{j}^{(t)} \rangle))$ rotation operation $\textbf{U}_{\textbf{R}}$ on it to obtain the state
\begin{equation}\label{eq:psi_4}
	| \psi_{4} \rangle = | \psi_{3} \rangle \otimes \left (\sqrt{1 - \langle u_{i}^{(s)} | u_{j}^{(t)} \rangle^{2}} | 0 \rangle^{R} + \langle u_{i}^{(s)} | u_{j}^{(t)} \rangle | 1 \rangle^{R} \right ).
\end{equation}

(5) Uncompute the $| w \rangle$, $| \lambda \rangle$ registers and measure the $R$ register to be $| 1 \rangle$. The final state
\begin{equation}\label{eq:psi_optimal_M} 
	| \psi_{M_{\ast}} \rangle = \sum_{i,j=1}^{d} \langle u_{i}^{(s)} | u_{j}^{(t)} \rangle | i \rangle | j \rangle, 
\end{equation}
which represents the matrix $M_{\ast} = P_{s}^{T} P_{t}$ can be obtained by ignoring the $B$, $C_{1}$, $C_{2}$ registers. 

For simplicity, the whole procedure above can be simplified as the unitary evolution
\begin{align}\label{eq:matrix multiplication}
	\textbf{U}_{\textbf{M}}(P_{s}, P_{t}) = \textbf{U}_{1}^{\dagger} \textbf{U}_{2}^{\dagger} \textbf{U}_{3}^{\dagger} \textbf{U}_{4} \textbf{U}_{3} \textbf{U}_{2} \textbf{U}_{1},
\end{align}
where
\begin{equation}\label{eq:U1U2U3U4}
	\begin{cases}
		\textbf{U}_{1} = \textbf{U}_{\textbf{P}}(P_{s}, P_{t})\textbf{U}_{2}, \\
		\textbf{U}_{2} = \textbf{I}^{I_{1}I_{2}C_{1}} \otimes \textbf{H}^{B}, \\
		\textbf{U}_{3} = \textbf{I}^{I_{1}I_{2}} \otimes \textbf{U}_{\textbf{PE}}^{BC_{1}C_{2}}(G), \\
		\textbf{U}_{4} = \textbf{I}^{I_{1}I_{2}BC_{1}} \otimes \textbf{U}_{\textbf{R}}^{C_{2}R}.
	\end{cases}
\end{equation}

Hence, the quantum state 
\begin{equation}\label{eq:hat_X_s}
	| \psi_{\hat{X}_{s}} \rangle = \textbf{U}_{\textbf{M}}(P_{s}, X_{s})| \psi_{0} \rangle = \sum_{i=1}^{n_{s}} | i \rangle | \hat{x}_{i}^{(s)} \rangle
\end{equation}
which contains the elements of the source domain subspace data $\hat{X}_{s}$. With the quantum states $| \psi_{M_{\ast}} \rangle$, $| \psi_{\hat{X}_{s}} \rangle$, the unitary operation $\textbf{U}_{\textbf{M}}(M_{\ast}, \hat{X}_{s})$ can be performed on $| \psi_{0} \rangle$ resulting in the quantum aligned source domain subspace state
\begin{equation}\label{eq:hat_X_a}
	| \psi_{\hat{X}_{a}} \rangle = \textbf{U}_{\textbf{M}}(M_{\ast}, \hat{X}_{s})| \psi_{0} \rangle = \sum_{i=1}^{n_{s}} | i \rangle | \hat{x}_{i}^{(a)} \rangle.
\end{equation}
Similarly, the quantum state corresponding to the target subspace data set $\hat{X}_{t}$ is
\begin{equation}\label{eq:hat_X_t}
	| \psi_{\hat{X}_{t}} \rangle = \textbf{U}_{\textbf{M}}(P_{t}, X_{t})| \psi_{0} \rangle = \sum_{j=1}^{n_{t}} | j \rangle | \hat{x}_{j}^{(t)} \rangle.
\end{equation}
In addition, the target aligned matrix $A$ can be represented by the quantum state
\begin{equation}\label{eq:psi_A}
	| \psi_{A} \rangle = \textbf{U}_{\textbf{M}}(\rho_{s}, \rho_{t}) | \psi_{0} \rangle.
\end{equation}

Therefore, the $D$-dimensional source and target domain data $X_{s}$, $X_{t}$ are projected to their corresponding $d$-dimensional subspace data $\hat{X}_{s}$, $\hat{X}_{t}$ respectively. The subspace coordinates of the two domains $P_{s}$, $P_{t}$ are subsequently aligned resulting in the transformation matrix $M$ and the aligned source domain data $\hat{X}_{a}$. Ultimately, the target labels $Y_{t}$ can be obtained by invoking a classifier. The pseudo-code of the QSA is presented in Algorithm~\ref{alg:QSA}.

\begin{algorithm}[t]
	\caption{Quantum subspace alignment}
	\KwIn{Source domain data $X_{s}$ with labels $Y_{s}$; target domain data $X_{t}$.}
	\KwOut{Target domain labels $Y_{t}$.}
	\emph{step 1}: Preprocess $X_{s}$, $X_{t}$ to obtain the corresponding principal components $P_{s}$, $P_{t}$ respectively by the qPCA. \\
	\emph{step 2}: Apply $\textbf{U}_{\textbf{M}}(P_{s}, P_{t})$ on $| \psi_{0} \rangle$ to obtain $| \psi_{M_{\ast}} \rangle$ representing the optimal transformation matrix $M_{\ast}$ as Eq.~\eqref{eq:psi_optimal_M}. \\
	\emph{step 3}: Apply $\textbf{U}_{\textbf{M}}(P_{s}, X_{s})$, $\textbf{U}_{\textbf{M}}(P_{t}, X_{t})$ on $| \psi_{0} \rangle$ respectively to obtain $| \psi_{\hat{X}_{s}} \rangle$, $| \psi_{\hat{X}_{t}} \rangle$  representing the source and target domain subspace data $\hat{X}_{s}$, $\hat{X}_{t}$ as in Eq.~\eqref{eq:hat_X_s}, Eq.~\eqref{eq:hat_X_t}. \\
	\emph{step 4}: Apply $\textbf{U}_{\textbf{M}}(M_{\ast}, \hat{X}_{s})$ on $| \psi_{0} \rangle$ to obtain $| \psi_{\hat{X}_{a}} \rangle$ representing the aligned source domain subspace data $\hat{X}_{a}$ as in Eq.~\eqref{eq:hat_X_a}. \\
	\emph{step 5}: Apply $\textbf{U}_{\textbf{M}}(\rho_{s}, \rho_{t})$ on $| \psi_{0} \rangle$ to obtain $| \psi_{A} \rangle$ representing the target aligned matrix $A$ as in Eq.~\eqref{eq:psi_A}. \\
	\emph{step 6}: Invoke a classifier to predict the target labels $Y_{t} = Classifier(\hat{X}_{a}, Y_{s}, \hat{X}_{t})$.
	\label{alg:QSA}
\end{algorithm}

\subsection{Algorithmic complexity}
\label{subsec:complexity}
To evaluate the performance of the classical and quantum SA, the algorithmic complexity of the two algorithms are analyzed as follows. 

In the data preprocessing, the classical SA utilizes the PCA to project the source and target domain data to the corresponding $d$-dimensional subspaces in time $O((n_{s}+n_{t})D + D^{3})$~\cite{pearson1901liii}. The procedure of aligning the two subspaces can be implemented in $O(D^{2} d)$~\cite{fernando2013unsupervised}. Compared with the classical SA, the source and target domain data sets are preprocessed by the qPCA in $O(d\log D)$~\cite{lloyd2014quantum}. Subsequently, the source domain subspace data $\hat{X}_{s}$ can be aligned to the target domain subspace data $\hat{X}_{t}$ in $O(\poly(\sqrt{D}))$~\cite{shao2018quantum}. Therefore, the procedure of DA can be implemented by the QSA presented in this paper with at least quadratic speedup. For the target label prediction, the local classifier, the quantum nearest neighbor algorithm~\cite{dang2018image}, can be selected as a candidate with the runtime in $O(\poly(\sqrt{n_{s}}))$ compared to the classical nearest neighbor algorithm in $O(n_{s} \log n_{s})$~\cite{gutin2002traveling}. In addition, the global classifier can also be invoked as an alternative resulting in exponential speedup in the number and dimension of the given data~\cite{rebentrost2014quantum}. Therefore, the subspace alignment can be achieved by the QSA with quadratic speedup in the whole procedure compared to the classical SA. 

\section{Variational quantum subspace alignment}
\label{sec:VQSA}
The VQSA is a common method to accomplish the subspace alignment with a variational hybrid quantum-classical procedure. Compared with the QSA presented in section~\ref{sec:QSA}, the VQSA can be achieved without high-depth quantum circuits and fully coherent evolution required by the QSA. In the first place, the given data are preprocessed by the PCA or the variational quantum state diagonalization algorithm~\cite{larose2018variational} resulting in $P_{s}$, $P_{t}$ as in the classical SA. Thus, the source and target domain subspace data $\hat{X}_{s} = P_{s}^{T}X_{s}$, $\hat{X}_{t} = P_{t}^{T}X_{t}$ can be obtained. Subsequently, the VQSA achieves the subspace alignment by designing parameterized quantum circuits to represent the data transformation and invoking a classical optimization algorithm to optimize the cost function. In the following, two different configurations of the VQSA are presented in section~\ref{subsec:E2E_VQSA} and section~\ref{subsec:MMB_VQSA} respectively.

\subsection{End-to-end VQSA}
\label{subsec:E2E_VQSA}
In the first configuration, we design an $L$-depth parameterized quantum circuit to represent the unitary evolution 
\begin{equation}
	\textbf{U}_{\theta^{(1)}} = \textbf{U}_{L}(\theta^{(1)}) \cdots \textbf{U}_{2}(\theta^{(1)}) \textbf{U}_{1}(\theta^{(1)})
	\label{eq:U_theta}
\end{equation}
with a set of parameters $\{ \theta^{(1)} \}$ to align the source domain subspace coordinates $P_{s}$ to the target domain subspace coordinates $P_{t}$. 

By minimizing the cost function
\begin{equation}
	L_1 = \left \vert \textbf{U}_{\theta^{(1)}} | P_{s}^{T} \rangle - | P_{t}^{T} \rangle \right \vert^{2}
	\label{eq:L_1}
\end{equation}
where $\textbf{U}_{\theta}$ represents an $L$-depth parameterized quantum circuit; $| P_{s}^{T} \rangle = \sum_{i=1}^{d} | u_{i}^{(s)} \rangle | i \rangle$, $| P_{t}^{T} \rangle = \sum_{j=1}^{d} | u_{j}^{(t)} \rangle | j \rangle$. The optimal transformation $\textbf{U}_{\theta^{(1)} \ast}$ can be subsequently applied to $\hat{X}_{s}$ to obtain the aligned source domain data $\hat{X}_{a} = \textbf{U}_{\theta^{(1)} \ast} \hat{X}_{s}$.

\subsection{Matrix-multiplication-based VQSA}
\label{subsec:MMB_VQSA}
In the second configuration, we design the ansatz states $| \psi_{M_{i}}(\theta^{(2)}) \rangle$ with a set of parameters $\{ \theta^{(2)} \}$ for $i = 1, \cdots, d$. Subsequently, the cost function
\begin{equation}
	L_{2} = \frac{1}{d} \sum_{i=1}^{d} \left \vert \frac{\langle \psi_{M_{i}}(\theta^{(2)}) | P_{s}^{T} | u_{i}^{(t)} \rangle}{\sqrt{\langle u_{i}^{(t)} | P_{s} P_{s}^{T} | u_{i}^{(t)} \rangle}} \right \vert^{2}
	\label{eq:L_2}
\end{equation}
is defined to be minimized to obtain the optimal ansatz states $| \psi_{M_{i}^{\ast}}(\theta^{(2)}) \rangle$ in the spirit of~\cite{bravo2019variational}. Hence, the optimal transformation matrix $M_{\ast} = \sum_{i=1}^{d} | \psi_{M_{i}^{\ast}}(\theta^{(2)}) \rangle \langle i |$ and the aligned source domain subspace data $\hat{X}_{a} = M_{\ast}^{T} \hat{X}_{s}$.

Having obtained the aligned source domain subspace data $\hat{X}_{a}$ and the target domain subspace data $\hat{X}_{t}$, we can invoke a classifier to predict the target labels $Y_{t}$. The pseudo-code of the VQSA is presented in Algorithm~\ref{alg:VQSA}.

\begin{algorithm}[tbt]
	\caption{Variational quantum subspace alignment}
	\KwIn{Source domain data $X_{s}$ with labels $Y_{s}$; target domain data $X_{t}$.}
	\KwOut{Target domain labels $Y_{t}$.}
	\emph{step 1}: Preprocess $X_{s}$, $X_{t}$ to obtain $P_{s}$, $P_{t}$ and $\hat{X}_{s}$, $\hat{X}_{t}$ by the PCA or the variational quantum state diagonalization algorithm. \\
	\emph{step 2}: Align $P_{s}$ to $P_{t}$ through the end-to-end VQSA in section~\ref{subsec:E2E_VQSA} or the matrix-multiplication-based VQSA in section~\ref{subsec:MMB_VQSA} resulting in the transformation matrix $M_{\ast}$ and the aligned source domain subspace data $\hat{X}_{a}$. \\
	\emph{step 3}: Invoke a classifier to predict the target labels $Y_{t} = Classifier(\hat{X}_{a}, Y_{s}, \hat{X}_{t})$.
	\label{alg:VQSA}
\end{algorithm}

\section{Numerical experiments}
\label{sec:numerical experiments}
\begin{table*}[htb]
\caption{\label{tab:table1}DA accuracies of all $4$ domain shifts on the synthetic data sets and the Iris data set}
\begin{ruledtabular}
\begin{tabular}{ccccc}
 &\multicolumn{2}{c}{Experiment 1}&\multicolumn{2}{c}{Experiment 2}\\
 & $D_{1} \rightarrow D_{2}$ & $D_{2} \rightarrow D_{1}$ & $D_{3} \rightarrow Iris$ & $Iris \rightarrow D_{3}$ \\ \hline
 NA & $50 \%$ & $49 \%$ & $33 \%$ & $50 \%$ \\
 Classical SA & $1 \%$ & $10 \%$ & $24 \%$ & $66 \%$\\
 VQSA & $\textbf{92 \%}$ & $\textbf{99 \%}$ & $\textbf{80 \%}$ & $\textbf{76 \%}$ \\
\end{tabular}
\end{ruledtabular}
\end{table*}

In this section, two different numerical experiments are presented to evaluate the performance of the VQSA compared with other models on different types of data sets. All the simulation experiments are implemented on a classical computer with the Python programming language and the Scikit-learn machine learning library~\cite{scikit-learn}. In addition, the source code and the selected parameters are accessible in Ref.~\cite{code}.

\subsection{Data sets}
\label{subsec:datasets}
Different types of data sets are selected to be applied to different tasks. In the first experiment, the two data sets $D_{1} \sim \mathcal{N}(\mu_{1}^{(1)} = \mu_{2}^{(1)} = 0, \sigma_{1}^{(1)} = \sigma_{2}^{(1)} = 1)$, $D_{2} \sim \mathcal{N}(\mu_{1}^{(2)} = \mu_{2}^{(2)} = 0, \sigma_{1}^{(2)} = \sigma_{2}^{(2)} = 2)$ are both generated from the normal distribution with the specified mean $\mu$ and variance $\sigma$. $D_{1}$, $D_{2}$ both contain $100$ six-dimensional data points distributed in two classes. For the second experiment, a synthetic data set $D_{3} \sim \mathcal{N}(\mu_{1}^{(3)} = \mu_{2}^{(3)} = \mu_{3}^{(3)} = 0, \sigma_{1}^{(3)} = \sigma_{2}^{(3)} = \sigma_{3}^{(3)} = 1)$ and the Iris data set~\cite{fisher1936use, anderson1936species} are selected. Concretely, the Iris data set contains $150$ four-dimensional samples distributed in three classes where each refers to a specific type of iris flower. The $D_{3}$ generated from the standard normal distribution has the same data dimension and feature space as the Iris data set but with a different distribution. In addition, the specified task $D_{A} \rightarrow D_{B}$ means that $D_{A}$ is selected as the source domain and $D_{B}$ is the target domain.

\subsection{Benchmark models}
\label{subsec:models}
The models selected in the numerical experiments are the no adaptation model (NA), the classical SA and the VQSA. The NA is a baseline model performed on the source and target domain data without any domain adaptation operations. The classical SA, the VQSA are performed on the source and target domain data sets respectively. The concrete structure of the parameterized quantum circuit in the VQSA is exactly the same as that in Ref.~\cite{he2020quantum}. In our work, the AdaGrad~\cite{duchi2011adaptive} is selected as the optimization algorithm.

\subsection{Results}
\label{subsec:results}
As presented in Table.~\ref{tab:table1}, all the three algorithms are performed on the two types of data sets with four domain shifts. In the first experiment, the NA model achieves $50 \%$ accuracy in the task $D_{1} \rightarrow D_{2}$ and $49 \%$ in the task $D_{2} \rightarrow D_{1}$. However, the classical SA performs unexpectedly poorly on the two tasks. For the VQSA, the model after the optimization is a $8$-layer parameterized quantum circuit. The performance of the VQSA is significantly better than other two algorithms. Thus, the VQSA shows superior data representation and transfer capabilities in the two specified tasks. In the second experiment, the classical SA achieves $66 \%$ accuracy in $Iris \rightarrow D_{3}$ better than $50 \%$ accuracy of the NA model. However, the $24 \%$ accuracy of the classical SA in the $D_{3} \rightarrow Iris$ is worse than the $33 \%$ accuracy of the NA demonstrating that the negative transfer appears in applying the classical SA to this task. Compared with the best performance of the other two algorithms, the VQSA can achieve at least $10 \%$ performance improvement. In the second experiment, the VQSA achieves accuracy higher than the other two models while effectively avoiding the negative transfer. The VQSA models in these two experiments are relative random parameterized quantum circuits. In our view, the performance of the VQSA can be promoted by further optimizing the circuit structures.

\section{Conclusion}
\label{sec:conclusion}
In this paper, we have presented two implementations of a representative DA algorithm, SA, on quantum devices. For the QSA, we perform the qPCA algorithm on the given raw data sets to obtain the subspace data sets with complexity logarithmic in the dimension of the data points in the data preprocessing section. Subsequently, the labelled subspace data are aligned to the target unlabelled subspace data based on the quantum basic linear algebra subroutines. Over the whole procedure, the QSA algorithm proposed in our work achieves at least quadratic speedup in algorithmic complexity compared with the classical SA algorithm. In addition, the VQSA is implemented on the near term quantum devices with a variational hybrid quantum-classical procedure. To evaluate the performance of the VQSA, we perform the VQSA and other models on the synthetic data sets and the Iris data set demonstrating that the VQSA can achieve competitive performance compared to the classical SA.

Aside form the two implementations provided in our work, some open questions still need to be further discussed. For the QSA, the quantum circuit depth of the QSA can be relatively high. In addition, the QSA requires fully coherent evolution and the qRAM which are prohibited at present. For the VQSA, how to design the quantum circuit to achieve optimal performance of the VQSA is another question. In general, although the algorithms proposed in this paper still have a lot to be improved, the algorithms in our work prove that quantum computation techniques can be applied to the field of transfer learning to accomplish machine learning tasks.   

\begin{acknowledgements}
	This work is supported by the National Key R\&D Program of China, Grant No. 2018YFA0306703.
\end{acknowledgements}

\nocite{*}

\bibliography{QSADA.bib}

\end{document}